%% file: main.tex
\documentclass{article}

\usepackage{arxiv}

\usepackage[utf8]{inputenc} 
\usepackage[T1]{fontenc}    
\usepackage{hyperref}       
\usepackage{url}            
\usepackage{booktabs}       
\usepackage{amsfonts}       
\usepackage{nicefrac}       
\usepackage{microtype}      
\usepackage{lipsum}\usepackage{amsmath,graphicx}
\usepackage{multirow}
\usepackage{xcolor}
\usepackage{csquotes}
\usepackage[caption=false]{subfig}
\usepackage[all]{xy}

\title{Multi-Contextual Design of Convolutional Neural Network for Steganalysis}

\author{
  Brijesh Singh \\
  \texttt{brijesh.singh@iitg.ac.in} \\
   \And
   Arijit Sur \\
 \texttt{arijit@iitg.ac.in} \\
  \And
 Pinaki Mitra \\
 \texttt{pinaki@iitg.ac.in} \\
}

\date{\vspace{-5ex}}

\begin{document}

\maketitle

\input{abstract}

\input{introduction}
\input{proposed}
\input{experiments}
\input{results}
\input{ablation}
\input{conclusions}

\bibliographystyle{unsrt}
\bibliography{refs}

\end{document}

%% file: abstract.tex
  In recent times, deep learning-based steganalysis classifiers became popular due to their state-of-the-art performance. Most deep steganalysis classifiers usually extract noise residuals using high-pass filters as preprocessing steps and feed them to their deep model for classification. It is observed that recent steganographic embedding does not always restrict their embedding in the high-frequency zone; instead, they distribute it as per embedding policy. Therefore, besides noise residual, learning the embedding zone is another challenging task. In this work, unlike the conventional approaches, the proposed model first extracts the noise residual using learned denoising kernels to boost the signal-to-noise ratio. After preprocessing, the sparse noise residuals are fed to a novel Multi-Contextual Convolutional Neural Network (M-CNET) that uses heterogeneous context size to learn the sparse and low-amplitude representation of noise residuals. The model performance is further improved by incorporating the Self-Attention module to focus on the areas prone to steganalytic embedding. A set of comprehensive experiments is performed to show the proposed scheme's efficacy over the prior arts. Besides, an ablation study is given to justify the contribution of various modules of the proposed architecture.

%% file: introduction.tex
\section{Introduction}
\label{sec:intro}
Image \textit{steganography} is the art of hiding a secret message within an image without compromising the visual quality as well as statistical undetectability. Steganography in digital images has been studied for decades. A variety of steganography can be found in \cite{steg1,steganography_book,steg2,li2011survey,gibbs,stc,suni,wow,hugo,mipod}.
On the other hand, \textit{steganalysis} is a science of detecting the traces of a hidden message within an innocuous-looking image. Steganalysis methods can be broadly categorized into two types based on feature extraction and classification, namely, handcrafted feature-based and deep feature-based methods. 

In general, it is assumed that the steganographic noise lies mainly in the high-frequency components of the stego image \cite{fridrich2011steganalysis}. Therefore, most steganalysis methods strive to extract the features from these components for better steganalytic detection. 
Pevny \textit{et al.}~\cite{spam} used a high-pass filter to compute the noise residual as the difference between the neighboring pixels. The noise residual is modeled using the Markov chain~\cite{markov}. The resulting transition probability matrix is used to train the Support Vector Machines (SVM)~\cite{svm} for steganalytic detection. Fridrich and Kodovsk\'{y}~\cite{srm} proposed the Spatial Rich Model (SRM), which uses various high-pass filters for feature extraction. These features are used to train the Ensemble classifier \cite{ensemble} for
steganalytic detection. The success of the handcrafted feature-based methods depends on the design of filters used for the feature extraction.

Recently, with the tremendous success of deep learning, researchers explored deep learning techniques to design detectors for steganalysis. Tan and Li~\cite{tan} presented a pre-trained Stacked Auto-Encoder \cite{SAE} based Convolutional Neural Network (CNN) model for steganalysis. Through this network, the authors tried to mimic the behavior of the SRM. However, the authors reported that the performance of the model could not match that of SRM \cite{srm}. Qian \textit{et al.}~\cite{gncnn} proposed a deep CNN-based model, namely \textit{GNCNN}. Unlike the traditional non-linearities, in GNCNN, Gaussian non-linearity is used as an activation function. The input images are first preprocessed using a fixed high-pass filter (KV\footnote{KV filter can be found in \cite{gncnn} in Eq. (2).} filter) to boost the signal-to-noise ratio (SNR). This exposure of noise is learned by the layers of the network for steganalytic detection. GNCNN achieved detection performance close to the SRM.
Xu \textit{et al.}~\cite{xunet} proposed a CNN architecture, \textit{Xunet}, which is equipped with the ABS layer to improve the statistical modeling of noise residual in subsequent layers, hyperbolic tangent (TanH) activation in early layers for capturing positive and negative embeddings, and $1\times1$ convolution in deeper layers for reducing the strength of modeling of steganalytic noise. Xunet also used the KV filter in the preprocessing steps to boost the SNR. The Xunet was the first model that achieved performance comparable to the SRM on S-UNIWARD~\cite{suni} and HILL~\cite{hill}. Ye \textit{et al.}~\cite{yenet} proposed the \textit{Yenet} model, which used thirty kernels initialized using linear and non-linear SRM~\cite{srm} filters to extract the diverse residuals. In addition, Yenet introduced a Truncated Linear Unit (TLU) to model the residual in a confined range in subsequent layers. Yedroudj \textit{et al.}~\cite{yedroudj} proposed a detector by fusing the most useful components used in recent detectors, such as XuNet and YeNet. Yedroudj-Net~\cite{yedroudj} also used SRM filters for the preprocessing of input images. Mehdi \textit{et al.}~\cite{srnet} proposed the \textit{SRNet} model. SRNet is the first end-to-end model, which does not include any preprocessing for residual computation. The authors reported that the first seven layers of the network compute the noise residual, and later five layers used these residuals for steganalytic detection. Singh \textit{et al.}~\cite{singh2019new} proposed a steganalysis scheme, which captured the noise residual with various size filters in each layer. The layers of the network are connected using skip connections for better modeling of the noise residuals. Singh \textit{et al.}~\cite{singh2020} devised a CNN architecture for steganalysis, which predicts the cover image from the given stego image. Then, the predicted cover image is subtracted from the stego image to compute noise residuals. The noise residual is learned by a classifier for steganalytic detection. The way this method trained the model offers only a single noise residual map, which is equivalent to using a single preprocessing filter. Singh \textit{et al.}~\cite{sfnet} utilized the FractalNet \cite{fractalnet} architecture to design an end-to-end steganalyzer called \textit{SFNet}, which maintains the balance between the width and the depth of the network for accurate detection.
You \textit{et al.}~\cite{siamese} used siamese-type CNN architecture to capture the relationships among sub-regions of the image for steganalytic detection.

The following observations are made from the literature: 
\begin{enumerate}
	\item A majority of the methods \cite{gncnn,xunet,yenet,yedroudj,singh2019new} use fixed high-pass filters in preprocessing stage for noise residual computation.
	\item The computed noise residuals are very low-amplitude signals that require a robust detector to be learned.
	\item The modern steganography algorithms distribute the embedding in the less predictable regions of the carrier image. Therefore, a mechanism is needed for the detectors to focus on these parts as well as the usual highly probable image regions.
\end{enumerate}
With the above motivation, our contributions in this paper are four-fold:
\begin{enumerate}
	\item A set of thirty filters is learned using a CNN model to replace the fixed high-pass filters used in preprocessing steps for residual computation from images~\cite{singh2020}.
	\item A multi-context design of a steganalyzer is devised to learn the low-amplitude and sparse noise residuals.
	\item The proposed model uses a Self-Attention~\cite{self-attention} mechanism to focus on areas that are likely to be affected by embeddings.
	\item Additionally, an ablation study is presented at the end for justification of proposed architecture.
\end{enumerate}
The rest of the paper is organized as follows: The proposed method is described in section \ref{sec:proposed}, the experimental study is discussed in section \ref{sec:experimental_study}, the experimental results are described in section \ref{sec:results}, an ablation study is presented in section \ref{sec:ablation}. Finally, this paper is concluded in section \ref{sec:conclusion}.

%% file: proposed.tex
\section{Proposed Method} 
\label{sec:proposed}
\begin{figure}[ht]
	\centering
	\subfloat[Cover]{
		\includegraphics[width=27mm]{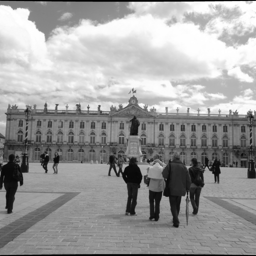}
	}
	\hspace{-2mm}
	\subfloat[WOW]{
		\includegraphics[width=27mm]{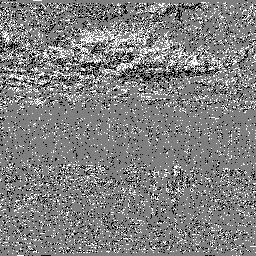}
	}
	\hspace{-2mm}
	\subfloat[S-Uniward]{
		\includegraphics[width=27mm]{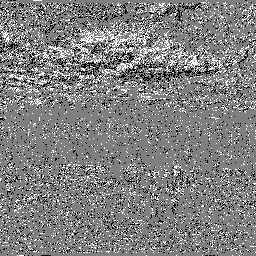}
	}
	\hspace{-2mm}
	\subfloat[HILL]{
		\includegraphics[width=27mm]{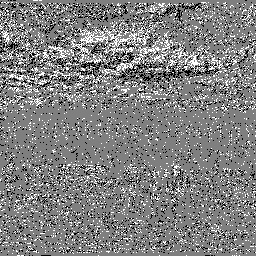}
	}
	\hspace{-2mm}
	\subfloat[MiPOD]{
		\includegraphics[width=27mm]{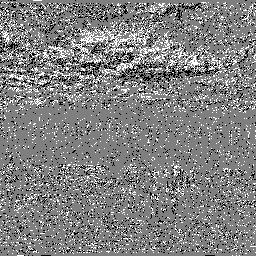}
	}
	\caption{Stego noise embedded by different steganographic algorithms - (b) WOW, (c) S-Uniward, and (d) HILL (e) MiPOD with payload = 0.4 bpp }
	\label{fig:noise_res}
\end{figure}
The name of the proposed model is \textit{M-CNet}, \enquote{Multi-contextual network for steganalysis.} In this model, we have used convolutions with different kernel sizes to get the responses having different contexts, hence the name \enquote{multi-context.} We believe that steganographic noise is not equally visible with a particular kernel. Instead, it (noise) can be traced more prominently with suitable kernel size depending on the local image statistics (context).
The proposed model can be roughly divided into two modules. The first module is an extension of our previous work~\cite{singh2020}, and the second module is a multi-context CNN model. 
\subsection{Rationale}
In principle, the modern content-adaptive steganography schemes hide most steganographic embedding in the noisy or texture region and relatively less in the smooth regions. The same observation can be made through Fig. \ref{fig:noise_res}, where (a) shows the cover image and (b)-(e) depicts the steganographic embedding done by WOW \cite{wow}, S-UNIWARD \cite{suni}, HILL \cite{hill}, and MiPOD \cite{mipod}, respectively. Noticeably, Fig. \ref{fig:noise_res} (b)-(e) shows that most steganographic embeddings are clustered towards the high texture regions and are sparsely distributed across the smooth areas.
This observation drives the steganalysis schemes to focus on the high-textured region to learn the features that may lead to better detection. Steganalysis methods \cite{spam,srm,gncnn,xunet,yenet,singh2019new} use different high-pass filters to suppress the image content and expose the noise content of an image. This allows the detectors to train on the noise domain rather than the image domain. However, this may not always be true. There are some recent embedding schemes that embed not only in the high-frequency zone but also in smoother areas~\cite{cmd1,cmd2}. For these cases, high-frequency-based filters may not be very useful.

With this motivation, we propose a denoiser subnetwork to compute the noise residual from a given image. For simplicity of reference, we call the denoiser subnetwork $\phi_{DN}$ subnetwork.
\subsection{Denoiser Subnetwork $\phi_{DN}$}
\label{denoiser_nw}
\begin{figure}[htb]
	\centering
	\includegraphics[width=0.7\linewidth]{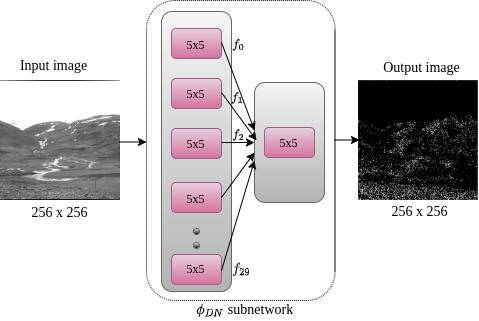}
	\caption{Architecture of the Denoiser subnetwork ($\phi_{DN}$).}
	\label{denoiser}
\end{figure}
\begin{figure}[htb]
	\centering
	\setlength{\tabcolsep}{1pt}
	\scalebox{0.67}{
		\begin{tabular}{cccccc}
			\bfseries image & \multicolumn{4}{c}{\textbf{corresponding ouput of $\phi_{DN}$}}\\
			\multirow{2}{*}{\includegraphics[width=2.5cm]{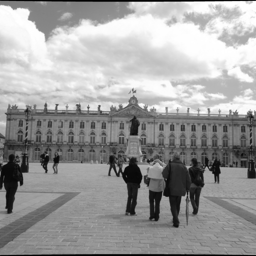}}&
			\includegraphics[width=2.5cm]{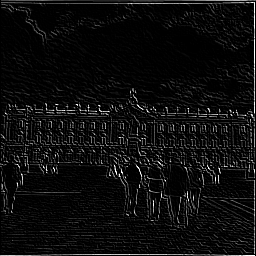}&
			\includegraphics[width=2.5cm]{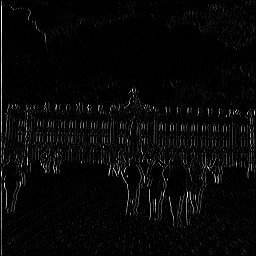}&
			\includegraphics[width=2.5cm]{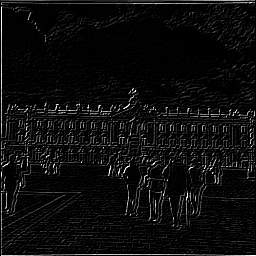}&
			\includegraphics[width=2.5cm]{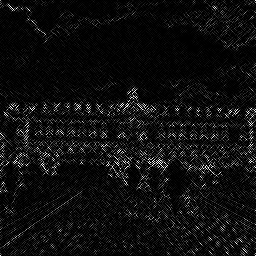}\\
			&
			\includegraphics[width=2.5cm]{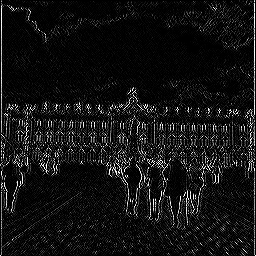}&
			\includegraphics[width=2.5cm]{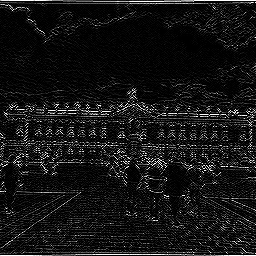}&
			\includegraphics[width=2.5cm]{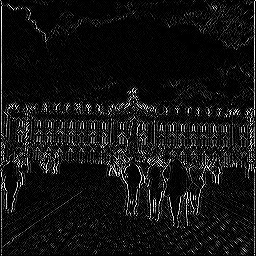}&
			\includegraphics[width=2.5cm]{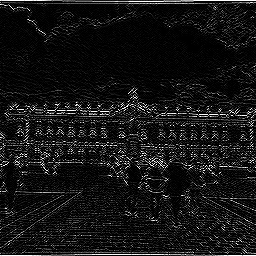}\\
			
			\multirow{2}{*}{\includegraphics[width=2.5cm]{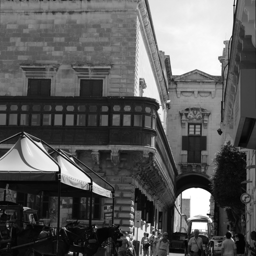}}&
			\includegraphics[width=2.5cm]{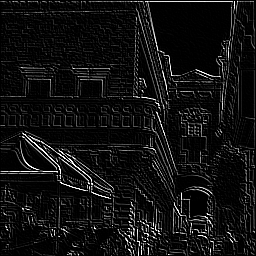}&
			\includegraphics[width=2.5cm]{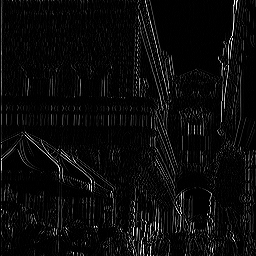}&
			\includegraphics[width=2.5cm]{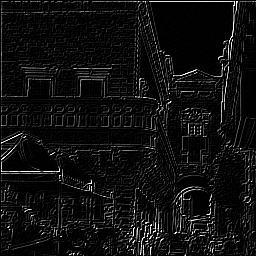}&
			\includegraphics[width=2.5cm]{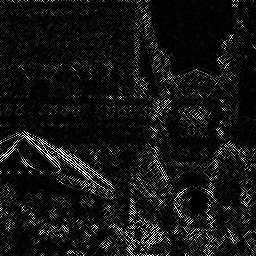}\\
			&
			\includegraphics[width=2.5cm]{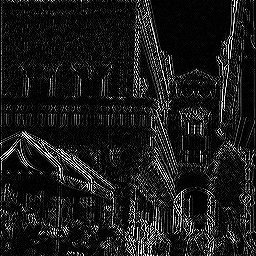}&
			\includegraphics[width=2.5cm]{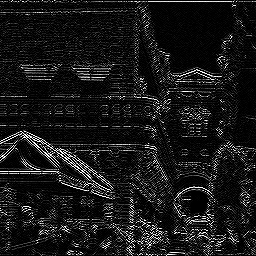}&
			\includegraphics[width=2.5cm]{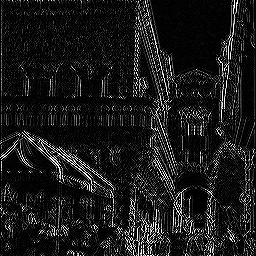}&
			\includegraphics[width=2.5cm]{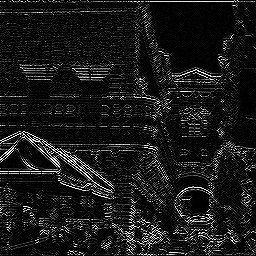}\\
	\end{tabular}}
	\caption{Sample output of $\phi_{DN}$ subnet: The first column represents input image, and columns 2 to 5 show the noise residual maps when preprocess using kernels of $\phi_{DN}$ subnet.}
	\label{fig:phi_dn_op} 
\end{figure}

The denoiser ($\phi_{DN}$) subnetwork extends one of our previous works \cite{singh2020}, where the cover image is predicted using a single-layered CNN from a stego image. Later, the noise residual is computed as the pixel-wise difference between the stego and the predicted cover images to train a classifier. We refer to the subnetwork mentioned above as \enquote{denoiser} due to its post-processing adaptability of predicting the stego noise residual. However, in this work, we train the $\phi_{DN}$ subnetwork to estimate the noise residual, bypassing the post-processing step directly. Furthermore, instead of utilizing 16 filters as in \cite{singh2020}, we propose to use 30 filters to extract the rich set of features that may capture the stego embeddings in both low and high-frequency zones. We have shown that such empirical changes in the $\phi_{DN}$ subnetwork's engineering may lead to a notable reduction in the detection error probability (see Section \ref{choice_of_denoising_kernels}). A graphical overview of the $\phi_{DN}$ subnetwork has been presented in Fig. \ref{denoiser}. Architecturally, the denoiser subnetwork consists of two convolution layers, with 30 and 1 filters, respectively. To formally define the regime of operations of $\phi_{DN}$ subnetwork, let $X$, $Y$ denote the cover and stego images, respectively. Consequently, the target cover and stego noise residuals can be written as $\mathcal{N}^c = | X - X |$ and $\mathcal{N}^s = | Y - X |$, respectively. Given an input image $X$ or $Y$, the proposed $\phi_{DN}$ subnetwork aims to estimate the corresponding $\mathcal{N}^c$ or $\mathcal{N}^s$ noise residual. The thirty kernels of the first convolution layer in the $\phi_{DN}$ subnetwork are initialized with SRM filters. It has been observed that such an initialization leads to faster convergence and a significant boost in detection accuracy. It should be mentioned that the pixel-wise formulation of noise residual may capture stego embeddings in both low and high-frequency zones of an image. However, unlike \cite{singh2020}, in this work, we utilize the inherent knowledge of the $\phi_{DN}$ subnetwork to train the proposed multi-contextual classifier. Precisely, instead of the predicted noise residual, we input the intermediate feature maps {$f_0, f_1, ... f_{29}$} generated by the 30 filters of the first layer in $\phi_{DN}$ to the M-CNet. By way of analysis, the intermediate feature representations of the final noise residual may have a variety of finer details (see Fig. \ref{fig:phi_dn_op}) that can be better leveraged by the multi-contextual filters of the proposed M-CNet. To the best of our knowledge, this is the first work that distills and incorporates the inherent information of the denoiser subnetwork in terms of intermediate feature representations and optimally trains the classifier.  

\subsection{Multi-context subnetwork}
\label{sec:multicontext}
\begin{figure*}[htb]
	\centering
	\includegraphics[width=0.995\linewidth]{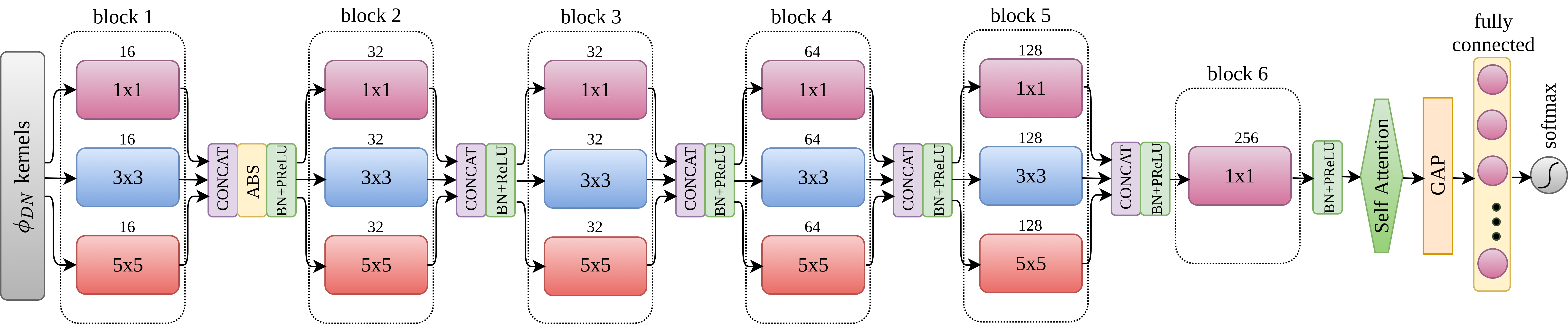}
	\caption{Architecture diagram of the proposed M-CNet.}
	\label{fig:architecture}
\end{figure*}
The purpose of the multi-context subnetwork is to learn the discriminative features that may not be learned using the fixed-size kernels due to the non-uniform sparsity of the stego noise in an image.
The multi-context subnetwork is combined with the preprocessing element $\phi_{DN}$ to form the M-CNet. 
As shown in Fig. \ref{fig:architecture}, the multi-context subnetwork comprises six convolution blocks, followed by a \textit{Self-Attention} layer~\cite{self-attention}, followed by a Global Average Pooling (GAP) layer, and a fully-connected layer followed by the softmax layer.
Each convolution block (\textit{block1} to \textit{block5}) is equipped with three different kernel sizes - $1\times1$, $3\times3$, and $5\times5$ to learn the noise residual at different contextual orientations. The learned features are then concatenated by using a \textit{CONCAT} layer before feeding to the next layer. After concatenation, the features are normalized using the \textit{Batch Normalization} (\textit{BN}) \cite{batch_norm} for the faster convergence of the network, followed by a Parametric ReLU (\textit{PReLU}) non-linearity \cite{he} before feeding to the next layer. The \textit{ABS} layer is applied to the concatenated features, which offers better learning of noise residual by preserving the negative as well as positive features~\cite{xunet}. The \textit{block6} is kept with a filter of size $1\times1$, which reduces modeling strength by keeping the salient features. The stego noise is fine-grained and sparse features scattered across the entire residual map. Therefore, a mechanism is needed
to learn these noises and focus on the areas that are likely to be most affected by embeddings. With this motivation, a \textit{Self-Attention} layer is used to enable the network to learn the features from the regions that are most likely to be affected by the steganographic embeddings. The employed Self-Attention mechanism adaptively refine the learned noise residual (output of the PReLU after \textit{block6}) by suppressing the irrelevant multi-contextual features. After the Self-Attention layer, the Global Average Pooling layer is used to reduce the dimensionality of the features by retaining only essential features. Following the \textit{GAP} layer, a fully connected network layer followed by the \textit{softmax} is used for binary classification using learned features. In order to come up with the final architecture, an extensive set of experiments have been conducted; the details are presented in section~\ref{sec:ablation}.

%% file: experiments.tex
\section{Experimental Details}
\label{sec:experimental_study}
This section presents the dataset details, steganography algorithms, and the evaluation metric used for training and evaluation of the models.
\subsection{Dataset}
\label{sec:dataset}
The performance of the proposed M-CNet is compared with SCA-Yenet~\cite{yenet} and SRNet~\cite{srnet}. Therefore, the same dataset configuration has been used for training, testing, and validation, with some minor changes. The entire dataset is consists of the union of BOSSbasee 1.01~\cite{bossbase} and BOWS-2~\cite{bows2}. Both of these datasets contain 10,000 grayscale images of size $512\times512$. Each of these images is first resized to $256\times256$ using the \textit{imresize} function of Matlab. The training set is comprised of 4,000 randomly selected images from the BOSSbase1.01 dataset and the entire BOWS2 images (10,000). The validation set consists of randomly selected 1,000 images from the remaining BOSSbase dataset and the rest 5,000 images are used for testing. Therefore, the training set contains $2\times14,000$ images (14,000 cover and 14,000 stego), the validation set includes $2\times1,000$ images, and the test set contain $2\times5,000$ images. Further, $2\times2,000$ pairs are randomly selected from the training set to train and validate ($2\times1,600$ for training, and $2\times400$ for validation) the $\phi_{DN}$ subnetwork. The stego image dataset is generated with S-Uniward~\cite{suni}, WOW~\cite{wow}, HILL~\cite{hill}, and MiPOD~\cite{mipod} steganography algorithms\footnote{The code for steganography algorithms are downloaded from {\color{magenta}\url{http://dde.binghamton.edu/download/stego\_algorithms/}}} using random keys.

All the models reported in this paper are trained on the same dataset splits, as reported in section \ref{sec:dataset}. The models are implemented using PyTorch~\cite{pytorch} on Nvidia V-100 with 32 GB GPU memory. 

\subsection{Training}
The proposed model is trained for the detection of spatial domain steganography schemes.\\
\subsubsection{Training $\phi_{DN}$ subnetwork} 
The $\phi_{DN}$ subnetwork is trained using $2\times1,600$ images and validated with $2\times400$ images. These images are not overlapped with any of the training, validation, and testing dataset of the proposed M-CNet model. 
The $\phi_{DN}$ is trained for 100 epochs, with a mini-batch of 20 images (10 cover and 10 stego), using the Adamax~\cite{adamax} optimizer. The learning rate is initialized to $10^{-3}$ and decayed every 25 epochs by a factor of $10^{-1}$. Each mini-batch images are subject to data augmentation, rotation by $90^{\circ}$, horizontal and vertical flipping, with a probability of 0.4.  The $\phi_{DN}$ subnetwork is trained by minimizing the \textit{mean-squared} loss ($\mathcal{L}_2$).
\begin{equation}
	\mathcal{L}_2=\|\hat{Y}-Y\|_2^2,
\end{equation}
where $Y$ and $\hat{Y}$ denote the input and the image estimated by $\phi_{DN}$ subnetwork, respectively. After the training, the model with the minimum validation error is chosen for the preprocessing of the proposed M-CNet model.

\subsubsection{Training M-CNet} 
The training of M-CNet is carried out on $2\times12,000$ training images. Adamax~\cite{adamax} optimizer is used with a minibatch of 20 images (10 cover and 10 stego) for 400 epochs. The learning rate is initialized to $10^{-3}$ and decayed every 40 epochs by a factor of $10^{-1}$. All the weights of convolutional kernels are initialized with \textit{Xavier initializer}~\cite{xavier}, and biases are initialized with 0. The weights of fully-connected neurons are initialized with Gaussian distribution with $\mu=0.0$ and $\sigma=0.01$.\\
Each mini-batch of training is subject to data augmentation by horizontal and vertical flipping and rotation by $90^{\circ}$ with a probability of 0.4.
The M-CNet is trained by minimizing the  binary cross-entropy loss $\mathcal{L}_{BCE}$.
\begin{equation}
	\mathcal{L}_{BCE}=-\frac{1}{N}\displaystyle\sum_{i=1}^N (y_i.log\hat{y}_{i}+(1-y_i).log(1-\hat{y}_i)),
\end{equation} 
where $N$ denotes the number of training examples, $y$, and $\hat{y}$ represent the true label and the predicted label, respectively.

\subsubsection{Evaluation metric} 
The performance of the models is evaluated using the minimum detection error probability under equal priors on the test set.
The minimum detection error probability is computed using the following formula:
\begin{equation}
	P_E = min_{P_{FA}}~\frac{1}{2}(P_{FA} + P_{MD}),
\end{equation}
where $P_{FA}$ and $P_{MD}$ denote the probabilities of false-alarm and missed-detection, respectively.
It is reported in the literature that one of the key requirements of a dependable steganalysis is the low false alarm rate \cite{dependable_steganalysis}. Therefore, recently, \textit{Weighted Area Under Curve} (WAUC) is also used for the evaluation of the dependability of the steganalysis methods \cite{alaska,siamese}. The WAUC gives more weight to the AUC below the true positive threshold than the area above the threshold, and then the area is normalized between 0 and 1. Specifically, first, the threshold is set to 0.4, then for the region above the threshold are given twice weight $\mathcal{W}=2$, and the region below the threshold are assigned weight $\mathcal{W}=1$.

In this paper, we have also given the ROC curve, Area Under ROC Curve (AUC), and the Weighted AUC (WAUC\footnote{code to compute WAUC is downloaded from: {\color{magenta}\url{https://www.kaggle.com/c/alaska2-image-steganalysis/overview/evaluation}}}) as an alternative metric of evaluation.
\par
The results presented in Section~\ref{sec:results} are reported for a random 50-50 split of the BOSSBase1.01 dataset.
However, to evaluate the detection performance across different splits, we trained the proposed M-CNet model on five different random 50-50 splits of BOSSBase, keeping BOWS2 fixed in training set on WOW with payload 0.4 bpp. As a result, the standard deviation of detection error ($P_E$) is obtained $\approx0.00359$ for these five splits.

%% file: results.tex
\section{Experimental Results and Discussion}
\label{sec:results}
In this section, the experimental results of the proposed M-CNet are presented and compared with the state-of-the-art detectors.
\begin{figure*}[ht]
	\centering
	\subfloat[WOW]{
		\includegraphics[width=67mm]{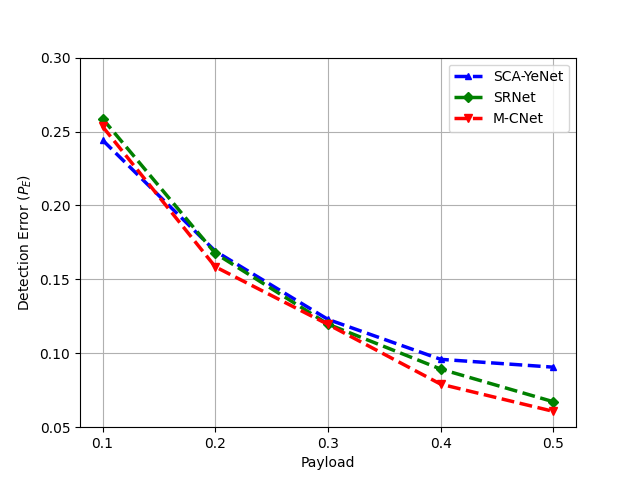}
	}
	\hspace{-3mm}
	\subfloat[S-Uniward]{
		\includegraphics[width=67mm]{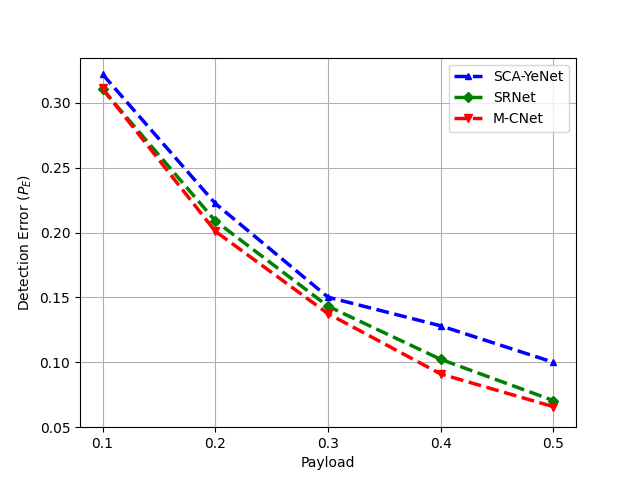}
	}
	\hspace{-3mm}
	\subfloat[HILL]{
		\includegraphics[width=67mm]{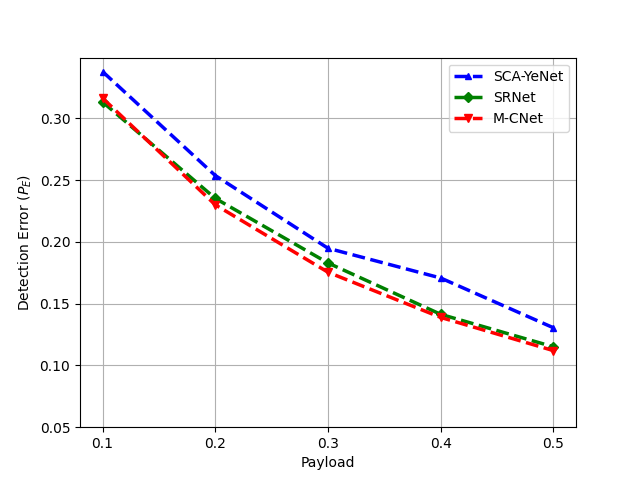}
	}
	\caption{Comparison among SCA-YeNet, SRNet, and M-CNet in terms of detection error probability $P_E$ on BOSSBase on (a) WOW, (b) S-Uniward, and (c) HILL steganography.}
	\label{fig:error_graph}
\end{figure*}

\begin{figure}[htb]
	\centering
	\includegraphics[width=0.6\linewidth]{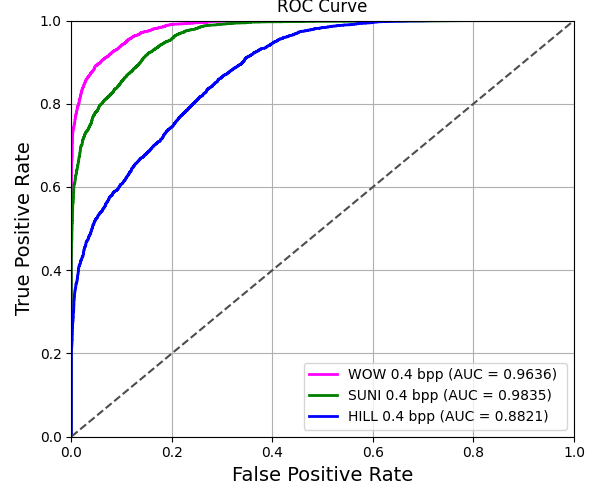}
	\caption{ROC curve of the proposed M-CNet when tested on WOW, S-UNIWARD, and HILL steganography with 0.4 bpp.}
	\label{fig:roc}
\end{figure}

The detection performance of the proposed M-CNet model is compared with the prior arts, namely, YeNet~\cite{yenet} and SRNet~\cite{srnet}. YeNet and SRNet are two spatial domain steganalytic detectors, which achieved state-of-the-art detection accuracy in the spatial domain. Both the methods are based on different training mechanisms. YeNet is based on the training with preprocessing filters, whereas SRNet is based on end-to-end training without preprocessing. The detection performance of these methods is compared with the proposed M-CNet while detecting the spatial domain steganography schemes, such as WOW~\cite{wow}, S-Uniward~\cite{suni}, and HILL~\cite{hill}. For a fair comparison with the proposed M-CNet, YeNet and SRNet are implemented with the experimental setup stated in the respective papers. The detection error probability $P_E$ of the steganalyzers is recorded in Table~\ref{tab:results}. The ROC curve for M-CNet is shown in Fig \ref{fig:roc}, and AUC/WAUC scores are given in Table \ref{tab:AUC_WAUC}. 
\begin{table}[ht]
	\caption{Detection error probability $P_E$ for M-CNet, YeNet, and SRNet. The best results are indicated in \textbf{bold face}.}
	\label{tab:results}
	\centering
	\begin{tabular}{ccccccl}
		\toprule
		\multirow{2}{*}{Embedding} & \multirow{2}{*}{Detector}&\multicolumn{5}{c}{Payload (bpp)}\\
		\cline{3-7}
		& & 0.1 & 0.2 & 0.3 & 0.4 & 0.5\\
		\midrule
		\multirow{3}{*}{WOW} & SCA-YeNet & \textbf{0.2442} & 0.1691 & 0.1229 & 0.0959 & 0.0906\\
		& SRNet & 0.2587 & 0.1676 & 0.1197 & 0.0893 & 0.0672\\
		& M-CNet & 0.2555 & \textbf{0.1649} & \textbf{0.1115} & \textbf{0.0759} & \textbf{0.0563}\\
		\midrule
		\multirow{3}{*}{S-UNI} & SCA-YeNet & 0.3220 & 0.2224 & 0.1502 & 0.1281 & 0.1000\\
		& SRNet & 0.3104 & 0.2090 & 0.1432 & 0.1023 & 0.0705\\
		& M-CNet & \textbf{0.3100} & \textbf{0.2010} & \textbf{0.1375} & \textbf{0.0910} & \textbf{0.0658}\\
		\midrule
		\multirow{3}{*}{HILL} & SCA-YeNet & 0.3380 & 0.2538 & 0.1949 & 0.1708 & 0.1305\\
		& SRNet & \textbf{0.3134} & 0.2353 & 0.1830 & 0.1414 & 0.1151\\
		& M-CNet & 0.3165 & \textbf{0.2301} & \textbf{0.1755} & \textbf{0.1389} & \textbf{0.1120}\\
		\bottomrule
	\end{tabular}
\end{table}
\begin{table}[ht]
	\caption{AUC and WAUC scores of the proposed M-CNet when detecting different types of steganography at differnt payloads.}
	\label{tab:AUC_WAUC}
	\centering
	\begin{tabular}{cccccl}
		\toprule
		\multirow{2}{*}{Embedding}&\multirow{2}{*}{Metric}&\multicolumn{4}{c}{Payload (bpp)}\\ 
		\cline{3-6} 
		& & 0.2 & 0.3 & 0.4 & 0.5\\
		\midrule
		\multirow{2}{*}{WOW}& AUC & 0.9324 & 0.9686 & 0.9835 & 0.9885\\
		&WAUC & 0.9517& 0.9776 & 0.9883 & 0.9917\\
		\midrule
		\multirow{2}{*}{SUNI}& AUC &0.8848 &0.9143 & 0.9636& 0.9830\\
		&WAUC &0.9170 & 0.9387 & 0.9740& 0.9879\\
		\midrule
		\multirow{2}{*}{HILL}& AUC &0 .8066& 0.8592&0.8821 &0.9066 \\
		&WAUC & 0.8425& 0.9343&0.9502 &0.9685\\
		\bottomrule
	\end{tabular}
\end{table}

\subsection{Curriculum Learning for training the M-CNet with lower payloads:}
It is a well-known fact that when a detector is trained on a steganography algorithm with a low payload, it performs worse, and sometimes it may not converge at all while training \cite{qian2016learning}. The solution to this kind of problem has been found using \textit{curriculum and transfer learning}~\cite{bengio2009curriculum,transfer_bengio}. Some of the recent steganalysis schemes \cite{yenet,srnet} used curriculum learning for training the model with the lower payloads. Following this trend, we trained the proposed M-CNet model with a higher payload (0.4 bpp) and then transferred its weights to train it for lower payloads (0.3 bpp) using a very small learning rate ($10^{-7}$), reducing every 100 epoch by a factor of $10^{-1}$. The proposed M-CNet model using the \textit{curriculum learning} is fine-tuned for 200 epochs. The proposed M-CNet model with \textit{curriculum learning} is evaluated for epochs between 51 to 200. The model with the best validation $P_E$ is evaluated, and results are presented in Table~\ref{tab:results}.

The results presented in Table~\ref{tab:results} and Fig.~\ref{fig:error_graph} show that the proposed M-CNet performed well when it is tested on WOW and S-Uniward while it attains the performance comparable to SRNet~\cite{srnet} on HILL embedding. It should also be noted that better detection performance is obtained for the embedding algorithms with higher payloads, such as 0.3 - 0.5 bpp. Further, the ROC curve presented in Fig~\ref{fig:roc} and AUC and WAUC (see Table~\ref{tab:AUC_WAUC}) as the alternate measure of detection performance. The ROC and AUC also depict that the proposed model performed better for the embedding algorithms WOW and S-uniward than HILL embedding. While the method such as SRNet design consideration is based on residual connections and end-to-end learning, the proposed M-CNet is explicitly designed to learn the sparse and low amplitude noise residual extracted by its preprocessing network($\phi_{DN}$). The experimental results show that the proposed model outperformed the prior arts. The detection efficacy of the proposed M-CNet model might be due to the multi-contextual design, which may have helped in learning the sparse and low amplitude noise residual spread over both low and high-frequency regions. Later, the adopted \textit{self-attention} mechanism might have helped in focusing on those regions more by suppressing the irrelevant regions.

Further, we evaluated the proposed M-CNet model with different architectural configurations. The details of the experiments are presented in section~\ref{sec:ablation}.

%% file: ablation.tex
\section{Ablation Study}
\label{sec:ablation}
This section presents a brief ablation study to assess the proposed model with different configurations and various scenarios. Unless explicitly specified, all the experiments presented in this section are performed on WOW steganography with a payload of 0.5 bpp.

\subsection{Configurations of Kernels in the $\phi_{DN}$}
\label{choice_of_denoising_kernels}
An obvious question one may ask, how do the different choices of $\phi_{DN}$ affect the performance of the proposed M-CNet model? Our goal with the $\phi_{DN}$ is to learn the kernels that can improve the noise residual computation. Therefore, we kept our search confined to a single layer. Table \ref{tab:choice_of_denoiser} presents the different configurations of filters explored for the $\phi_{DN}$. The first column represents the number of filters ($n$), the second and the third column represent the filter size ($s$) used and the corresponding detection error probability $P_E$, respectively. The best $P_E$ is achieved in this experiment for $n=30$ and $32$, and $s=3\times3$. These results are obtained when all the kernels are initialized with the \textit{Kaiming initialization}~\cite{he}. We also investigated the initialization of the kernels using SRM, which has kernel sizes of $1\times1$ to $5\times5$. The smaller size SRM kernels are padded to zeroes to match the dimension of $5\times5$. In order to initialize with the SRM kernels, we kept the no. of filters $n=30$ and kernel size $s=5\times5$. The details of initialization are presented in section \ref{sec:denoiser_initialization}. We also explored a few smaller architectures with a few layers for the $\phi_{DN}$, but could not get any promising results.
\begin{table}[ht]
	\caption{Detection error probability ($P_E$) with different filter size and no. of filters in $\phi_{DN}$.}
	\label{tab:choice_of_denoiser}
	\centering
	\begin{tabular}{ccl}
		\toprule
		\# filters ($n$) & filter size ($s$) & $P_E$ \\
		\midrule
		\multirow{2}{*}{16} & $3\times3$ & 0.0820\\
		& $5\times5$ & 0.0873\\
		\midrule
		\multirow{2}{*}{30} & $3\times3$ & \textbf{0.0800}\\
		& $5\times5$ & 0.0810\\
		\midrule
		\multirow{2}{*}{32} & $3\times3$ & 0.0811\\	
		& $5\times5$ & 0.0830\\
		\midrule
		\multirow{2}{*}{64} & $3\times3$ & 0.0895\\
		& $5\times5$ & 0.0937\\
		\bottomrule
	\end{tabular}
\end{table}
\subsection{Kaiming vs. SRM vs. Gabor initialization}
\label{sec:denoiser_initialization} 
It has been widely established that the learning-based models may achieve poor convergence when initialized with the random weights derived from the Gaussian distribution \cite{xavier}. To overcome this drawback, we experimented by initializing the kernels of the $\phi_{DN}$ module with different filters, namely, \textit{Kaiming} \cite{he}, \textit{SRM} filters \cite{srm}, and \textit{Gabor} filters \cite{gabor}. For our experiment, $30$ \textit{Gabor} filter are obtained with following parameters: Scale $\sigma\in\{0.5,1\}$, Wavelength of the cosine function $\lambda=\sigma/0.56$, Spatial aspect ratio $\gamma=0.5$, and fifteen orientations $\theta\in[0,14\pi/15]$. The experimental results are tabulated in Table \ref{tab:hpf_ablation}. The \textit{SRM}, and \textit{Gabor} Kernels are first padded with zeroes to match the dimension to $5\times5$ before the initialization. The $\phi_{DN}$ converged faster while training when initialized with \textit{SRM} kernels. We also evaluated the performance of the proposed M-CNet model with these initializations; results are shown in Table~\ref{tab:initialization_of_denoiser}. The results show that the proposed M-CNet model also performed better when initialized with \textit{SRM} kernels.
\begin{table}[ht]
	\caption{Detection error probability when the $\phi_{DN}$ subnetwork is initialized with different types of kernels.}
	\label{tab:initialization_of_denoiser}
	\centering
	\begin{tabular}{cccl}
		\toprule
		Initialization & Kaiming & SRM & Gabor\\
		\midrule
		$P_E$ & 0.0810 &\textbf{0.0563} & 0.0723\\
		\bottomrule	
	\end{tabular}
\end{table}

\subsection{Split vs. end-to-end training of the $\phi_{DN}$ subnetwork}
The training of the proposed M-CNet model is carried out in two phases (we call it split training). In the first phase, the $\phi_{DN}$ subnetwork is trained.
In the second phase, the learned weights of the $\phi_{DN}$ are used in preprocessing, and the entire network is trained, keeping the learned kernel weights of the $\phi_{DN}$ fixed. Nevertheless, to investigate the performance of the proposed M-CNet model when trained in the end-to-end fashion, both the networks ($\phi_{DN}$ and multi-context subnetworks) are clubbed together, keeping the $\phi_{DN}$ in the front end of the M-CNet. As a result, the error detection probability $P_E$ is found to be $0.0848$ for end-to-end trained M-CNet, which is inferior to the performance with the split training (\textbf{$0.0563$}).


\subsection{Detection performance of the proposed M-CNet when the $\phi_{DN}$ subnetwork is replaced with handcrafted filters like- SRM, KV, or Gabor}
\label{sec:srm_kv_gabor}

To this end, we trained the model without any preprocessing and with preprocessing using KV, SRM, and Gabor filters \cite{gabor}. The KV filter is a single $5\times5$ filter, which has been used in numerous steganalyzers~\cite{gncnn,xunet} for preprocessing. The SRM~\cite{srm} filter bank consists of 30 linear and non-linear filters. All these kernels are resized to $5\times5$ before using with M-CNet. A set of 2D Gabor filters \cite{gabor} has been used by Li \textit{et al.}~\cite{restnet} in the preprocessing stage of the detector for better extraction of noise residual. The Gabor filters are obtained using the setting stated in section \ref{sec:denoiser_initialization}.
\begin{table}[ht]
	\renewcommand{\arraystretch}{1.3}
	\caption{Detection error probability when different kind of filters are used in preprocessing stage of the proposed M-CNet.}
	\label{tab:hpf_ablation}
	\centering
	\begin{tabular}{ll}
		\toprule
		Preprocessing using & $P_E$\\
		\midrule
		No filter & 0.1115\\
		KV filter  & 0.1079\\
		Gabor filter & 0.0959\\
		SRM filters & 0.0824\\
		$\phi_{DN}$ kernels & \textbf{0.0563}\\
		\bottomrule		
	\end{tabular}
\end{table}
The best detection error probability is found when the model is trained with preprocessing using the $\phi_{DN}$ kernels. The result shows that the preprocessing elements, which are learned, are comparatively better than the fixed ones.

\begin{figure*}[htb]
	\centering
	\setlength{\tabcolsep}{1pt}
	\scalebox{0.52}{
		\begin{tabular}{cccccccccc}
			\bfseries image & \bfseries Stego noise & \multicolumn{4}{c}{\bfseries ------- Feature maps predicted by \textit{Self-Attention} layer -------}&\multicolumn{4}{c}{\bfseries ----------------- Feature maps predicted by \textit{block6} -----------------}\\
			
			\includegraphics[width=2.5cm]{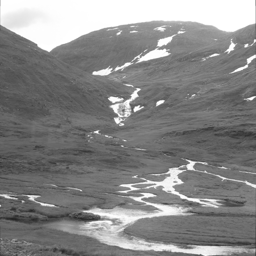}&
			\includegraphics[width=2.5cm]{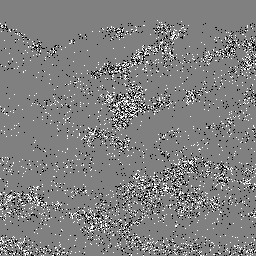}&
			\includegraphics[width=2.5cm]{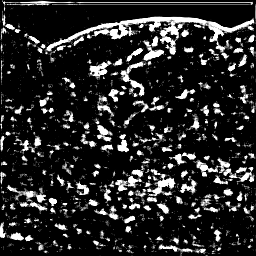}&
			\includegraphics[width=2.5cm]{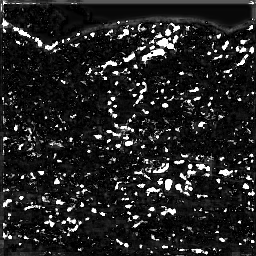}&
			\includegraphics[width=2.5cm]{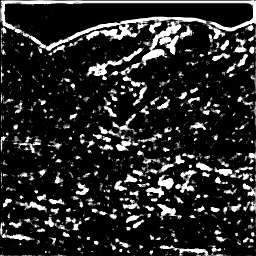}&
			\includegraphics[width=2.5cm]{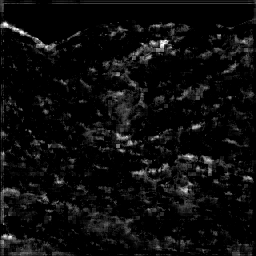}&
			\includegraphics[width=2.5cm]{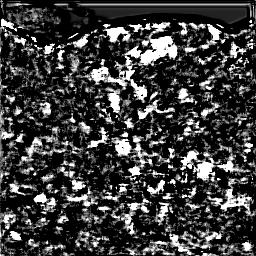}&
			\includegraphics[width=2.5cm]{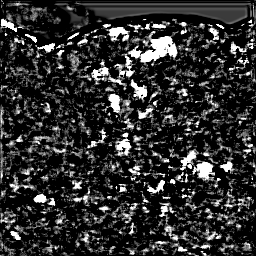}&
			\includegraphics[width=2.5cm]{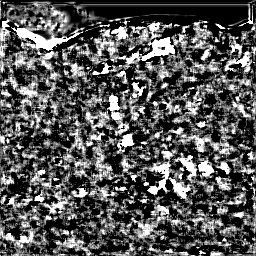}&
			\includegraphics[width=2.5cm]{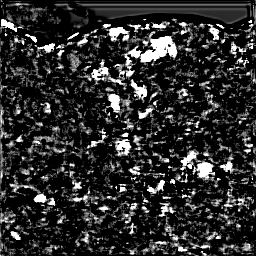}\\			
			\includegraphics[width=2.5cm]{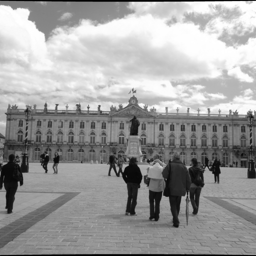}&
			\includegraphics[width=2.5cm]{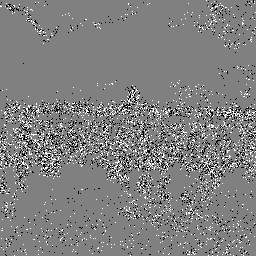}&
			\includegraphics[width=2.5cm]{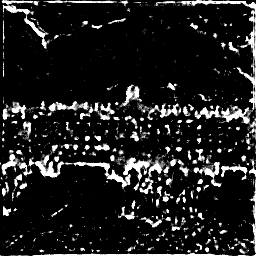}&
			\includegraphics[width=2.5cm]{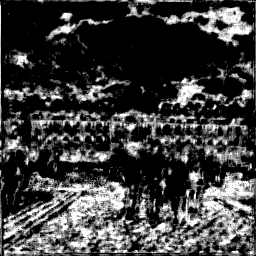}&
			\includegraphics[width=2.5cm]{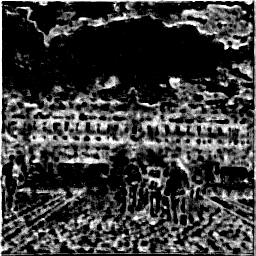}&
			\includegraphics[width=2.5cm]{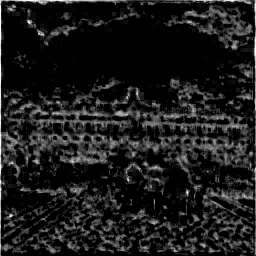}&
			\includegraphics[width=2.5cm]{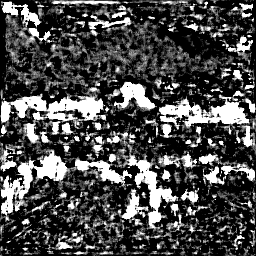}&
			\includegraphics[width=2.5cm]{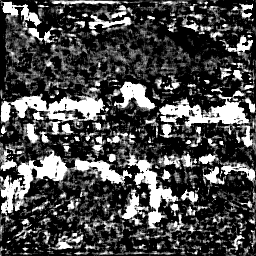}&
			\includegraphics[width=2.5cm]{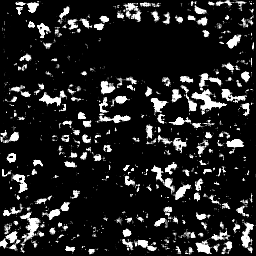}&
			\includegraphics[width=2.5cm]{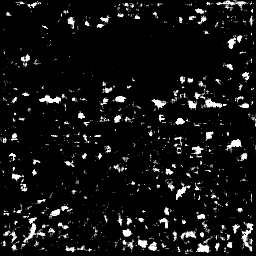}\\
			
			\includegraphics[width=2.5cm]{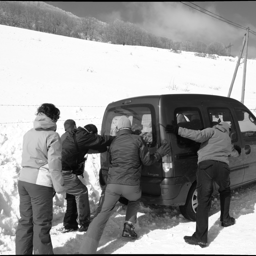}&
			\includegraphics[width=2.5cm]{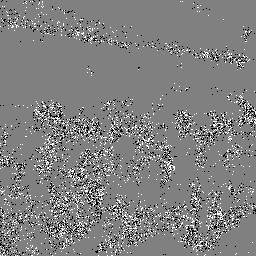}&
			\includegraphics[width=2.5cm]{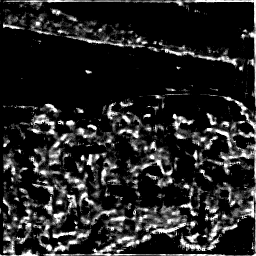}&
			\includegraphics[width=2.5cm]{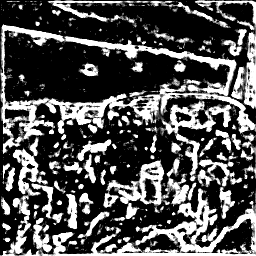}&
			\includegraphics[width=2.5cm]{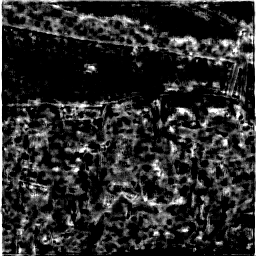}&
			\includegraphics[width=2.5cm]{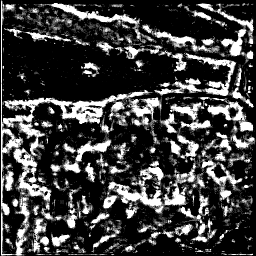}&
			\includegraphics[width=2.5cm]{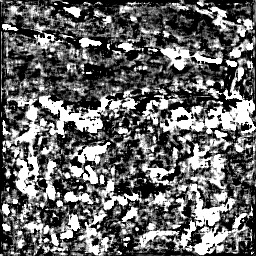}&
			\includegraphics[width=2.5cm]{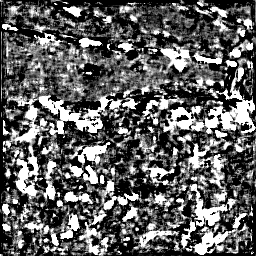}&
			\includegraphics[width=2.5cm]{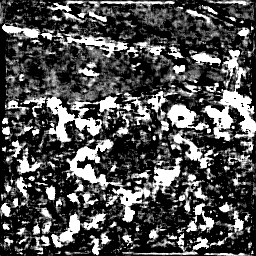}&
			\includegraphics[width=2.5cm]{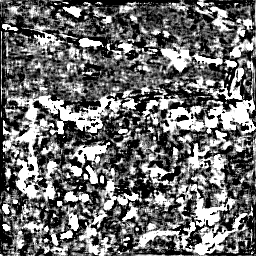}\\
			
			\includegraphics[width=2.5cm]{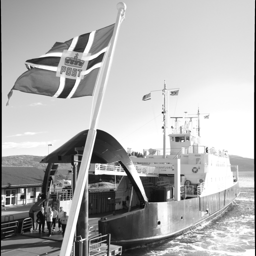}&
			\includegraphics[width=2.5cm]{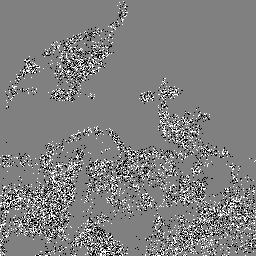}&
			\includegraphics[width=2.5cm]{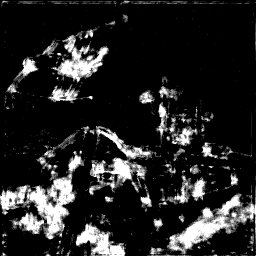}&
			\includegraphics[width=2.5cm]{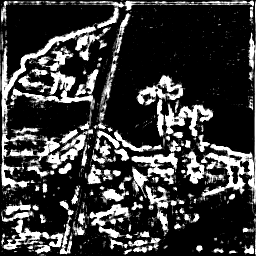}&
			\includegraphics[width=2.5cm]{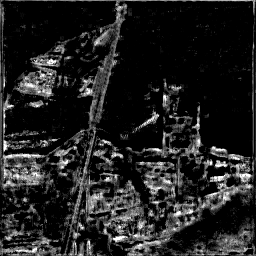}&
			\includegraphics[width=2.5cm]{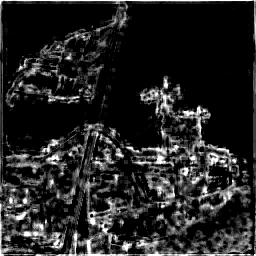}&
			\includegraphics[width=2.5cm]{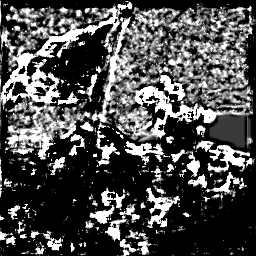}&
			\includegraphics[width=2.5cm]{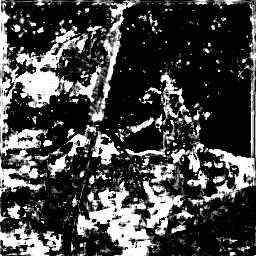}&
			\includegraphics[width=2.5cm]{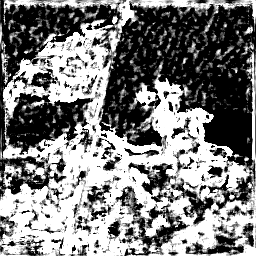}&
			\includegraphics[width=2.5cm]{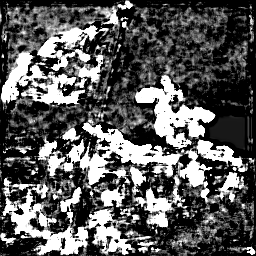}\\
	\end{tabular}}
	\caption{Sample: The first two columns represent an image and corresponding noise residual ($\mathbf{N}=\mathbf{Y}-\mathbf{X}$).
		The next four columns (col. 3-6) show the 4 feature maps (out of 256) predicted by the \textit{Self-Attention} layer of M-CNet, and the col. 7-10 show the 4 feature maps (out of 256) predicted by \textit{block6} (M-CNet without \textit{Self-Attention}).}
	\label{fig:attention_maps} 
\end{figure*}
\subsection{Choice of filter configuration in multi-context subnetwork}

We carried out experiments with different configurations of the proposed M-CNet, keeping only one type of kernel in each convolutional layer except for the last layer where 1x1 kernel size is used. The results of this experiment are presented in Table \ref{tab:filter_size}. When all the layers of the proposed M-CNet are equipped with the $1\times1$ kernel size, the detection error is found to be 0.1319, which might be due to the smaller context size alone is not suitable for capturing the stego features. The better detection error 0.0968 is obtained for the model with kernel size $3\times3$. Nevertheless, considering the low amplitude and sparse characteristics of stego noise, any single type of kernel alone may not be suitable to capture such features.
Therefore, we further explored the models with multiple context sizes in each layer to capture these features more precisely. The results of this experiment are presented in Table \ref{tab:types_of_filter_per_layer}. We initially experimented with combinations of two different size kernels in each convolution layer, keeping the last layer fixed with only 1x1 kernels. Finally, we experimented with all three different kernel sizes in each layer. The model with this configuration performed better than the former combinations.
\begin{table}[ht]
	\caption{Detection error probability when only one type of filter size is used in each layer of the multicontext network.}
	\label{tab:filter_size}
	\centering
	\begin{tabular}{ll}
		\toprule
		Layers configuration & $P_E$\\
		\midrule
		All layers $1\times1$ & 0.1319\\
		All layers $3\times3$ & \textbf{0.0968}\\
		All layers $5\times5$ & 0.1010\\
		First two layers $5\times5$$\rightarrow$ three layers $3\times3$ & 0.1054\\
		\bottomrule		
	\end{tabular}
\end{table}

\begin{table}[ht]
	\caption{Detection error probability $P_E$ when multiple filters of different size are used in each layer of the network.}
	\label{tab:types_of_filter_per_layer}
	\centering
	\begin{tabular}{ll}
		\toprule
		Layers configuration & $P_E$\\
		\midrule
		$1\times1$ and $3\times3$ & 0.0939 \\
		$1\times1$ and $5\times5$ & 0.0872\\
		$3\times3$ and $5\times5$ & 0.0786\\
		$1\times1$, $3\times3$, and $5\times5$ & \textbf{0.0563}\\
		\bottomrule		
	\end{tabular}
\end{table}

\subsection{Increasing and Decreasing depth and no. of filters per layer of the proposed M-CNet model}
An extensive set of experiments have been conducted to develop the final architecture of the M-CNet; some of them are presented here. The first question is, how does the performance changes along with the depth of the network? To this end, we conducted the experiments keeping all the components fixed and varying only the depth ($d$ = no. of blocks) from 2 to 8. Beyond $d=8$, the model exhausted the available resources for computation. The result presented in Table~\ref{tab:depth_variation} shows that the M-CNet performed best for $d=6$ and $7$. However, we kept the configuration of the M-CNet model to $d=6$ since it saves the number of trainable parameters and lowers the training time. It can infer from the results that the model with $d=2$ is very weak to learn the stego embedding, whereas, for  $d=8$, the low-amplitude stego noise may vanish along with the depth of the network.
\begin{table}[h]
	\caption{Detection error $P_E$ with variation in depth of M-CNet.}
	\label{tab:depth_variation}
	\centering
	\begin{tabular}{cccccccl}
		\toprule
		Depth ($d$) & 2 & 3 & 4 & 5 & 6 & 7 & 8\\
		\midrule
		$P_E$ &0.2239&0.1080&0.0980&0.0729&\textbf{0.0563}&0.0610&0.0759\\
		\bottomrule
	\end{tabular}
\end{table}

\subsection{Choice of activation}
There are various choices for activation function such as \textit{Sigmoid}, \textit{TanH}, \textit{ReLU}, Leaky ReLU (\textit{LReLU}), Parametric ReLU (\textit{PReLU}). 
\begin{equation*}
	Sigmoid(\theta_i)=\frac{1}{1+e^{-\theta_i}};~~~~TanH(\theta_i)=\frac{e^{\theta_i}-e^{-\theta_i}}{e^{\theta_i}+e^{-\theta_i}},
\end{equation*}
\begin{equation*}
	f(\theta_i)= \begin{cases}
		\theta_i, & \text{if } \theta_i> 0\\
		\alpha_i.\theta_i, & \text{if } \theta_i\leq 0,
	\end{cases}
\end{equation*}
where $\theta_i$ denote the input to the nonlinear activation on the $i^{th}$ channel. $f(\theta_i)$ denote the \textit{ReLU} activation when the parameter $\alpha_i=0$, \textit{LReLU} when $\alpha_i$ is fixed to a small constant ($\alpha_i= 0.01$), and \textit{PReLU} when $\alpha_i$ is learned with the model.

The aim of \textit{LReLU} is to avoid the zero gradients that occur in \textit{ReLU}. But, experiments in \cite{LRelu} showed that replacing the \textit{ReLU} with the \textit{LReLU} has a very negligible impact on accuracy. However, \textit{PReLU} adaptively learns the parameters jointly with the model and also has a considerable impact on the accuracy \cite{he}. For steganalytic detectors, \textit{PReLU} helps to learn the positive as well negative embedding more precisely by retaining the negative values using learned parameters, which might be lost when \textit{ReLU} is used. We experimented with \textit{Sigmoid}, \textit{TanH}, \textit{ReLU}, and \textit{PReLU} activation to compare the performance when different types of activation are used. The $P_E$ for each of these activations is recorded in Table \ref{tab:activation}. Experimentally, the best detection error is observed when \textit{PReLU} activation is used throughout the network.
\begin{table}[ht]
	\caption{Detection error $P_E$ with variation in depth of M-CNet.}
	\label{tab:activation}
	\centering
	\begin{tabular}{cccccl}
		\toprule
		Activation & Sigmoid & TanH & TanH$\rightarrow$ReLU & ReLU & PReLU\\
		\midrule
		$P_E$ &0.1811 &0.0776&0.0670 &0.0607&\bfseries0.0563\\
		\bottomrule
	\end{tabular}
\end{table}
\subsection{With or without Self-Attention?}

To answer this question, we trained the M-CNet without the \textit{Self-Attention}~\cite{self-attention} and compared the performance with the M-CNet. The detection error is found to be $P_E=0.0855$. A set of samples for the model with and without the Self-Attention are presented in Fig.~\ref{fig:attention_maps}. In Fig. \ref{fig:attention_maps}, col. 1 shows a cover image, col. 2 shows the noise embedding ($\mathbf{N}=\mathbf{Y}-\mathbf{X}$) by WOW steganography with payload 0.5 bpp, the col. 3-6 represent the features predicted by the Self-Attention layer of the M-CNet, and col. 7-10 represent the features predicted by \textit{block 6} when the Self-Attention layer is not used in M-CNet. The samples show that the outputs predicted by the layers provide a variety of features by assigning high values ($255$) to the highly textured and low values ($0$) to the smooth regions. Nevertheless, it can be clearly observed that the output maps predicted by Self-Attention focus more on the regions that contain steganographic embeddings. This precise feature learning of Self-Attention caused the increase in detection accuracy by $\approx 3\%$.

\subsection{Detection performance of the proposed M-CNet for stego-source mismatch}
To this end, we trained the proposed M-CNet model on one steganography algorithm and tested it with another with the same payload. The results presented in Table~\ref{tab:stego_source_mismatch} show that when the model is trained on the easy to detect algorithm WOW and tested on the hard-to-detect algorithm MiPOD, the model performed worse. Comparatively, it performed better when trained on MiPOD and tested on WOW and several other steganography.
\begin{table}[ht]
	\caption{Detection error probability $P_E$ of the proposed M-CNet for stego-source mismatch scenario on different embedding with payload 0.4 bpp}
	\label{tab:stego_source_mismatch}
	\centering
	\begin{tabular}{ccccl}
		\toprule
		Train\textbackslash Test & WOW & S-UNI & HILL & MiPOD \\
		\midrule
		WOW & 0.0759 & 0.1479 & 0.3301 & 0.2850\\
		S-UNI & 0.1082 & 0.0910 & 0.2501 & 0.2095\\
		HILL & 0.1629 & 0.2755 & 0.1389 & 0.2160\\
		MiPOD & 0.1412 & 0.1580 & 0.1812 &  0.1441\\
		\bottomrule
	\end{tabular}
\end{table}

\subsection{Detection performance of the proposed M-CNet for cover-source mismatch}
\begin{table}[htb]
	\caption{Detection error $P_E$ for proposed model cover-source mismatch for WOW 0.4 bpp}
	\label{tab:cover_source_mismatch}
	\centering
	\begin{tabular}{ccl}
		\toprule
		Trained on & Tested on & $P_E$\\
		\midrule
		Imagenet & BOSSbase & 0.3936\\
		BOSSbase $\cup$ BOWS2& Imagenet & 0.4725\\
		\bottomrule
	\end{tabular}
\end{table}
In order to assess the performance of the M-CNet for the scenario when training images are coming from one distribution and test images from another with the same embedding and payload (WOW 0.4 bpp). We performed the following set of experiments: (1) Trained on the training set explained in section \ref{sec:dataset} (here, we referred to it as BOSSbase $\cup$ BOWS2) and tested on $5,000$ randomly selected images from \textit{Imagenet}\footnote{Imagenet images are first converted to grayscale using \textit{rgb2gray} and then resized to $256\times256$ using \textit{imresize} function of Matlab before the embedding.} \cite{imagenet}.
(2) Trained on 10,000 real-world images from the \textit{Imagenet} dataset and tested on the BOSSbase test set.
Since the \textit{Imagenet} dataset consists of a wide variety of images, when the M-CNet is trained on \textit{Imagenet} and tested on BOSSbase performed better than when it is trained on BOSSbase $\cup$ BOWS2 and tested on Imagenet. The results of this experiment are tabulated in Table \ref{tab:cover_source_mismatch}.

%% file: conclusions.tex
\section{Conclusions}
\label{sec:conclusion}
In this paper, a steganalytic detector, M-CNet, is proposed to learn the feature for the steganographic embeddings with multiple context sizes. The proposed M-CNet model is equipped with learned kernels for preprocessing, which offers diverse noise residual by exposing noise components and suppressing image components. Further, the M-CNet employed the \textit{Self-Attention} mechanism to focus on the image regions, which are likely to be more affected by steganographic embeddings. Through an ablation study, the design of the proposed model is justified in favour of its detection performance. The experimental results show that the M-CNet outperformed the state-of-the-art detectors.